\documentclass[10pt]{article}

\usepackage{savesym}
\savesymbol{iint}
\usepackage{txfonts}
\restoresymbol{TXF}{iint}
\usepackage{mathptmx}
\usepackage{paralist}
\usepackage[tiny,compact]{titlesec}
\usepackage{amssymb}
\usepackage{paralist}
\usepackage{graphicx}
\usepackage{url}
\usepackage{epsfig,amsfonts,latexsym}

\newenvironment{definition}[1][Definition]
{\begin{trivlist}
\item[\hskip \labelsep {\bfseries #1}]}{\end{trivlist}}

\newcommand{\move}[1]{\stackrel{#1}{\longrightarrow}}

\begin{document}

\title{A Formal Verification Technique for Architecture-based Embedded Systems in EAST-ADL}



\author{Eun-Young Kang\\
The Maersk Mc-Kinney Moller Institute\\
The University of Southern Denmark, \\
Odense, Denmark\\
eyk@mmmi.sdu.dk}

%

\maketitle

\begin{abstract}
Development of quality assured software-intensive systems, such as
automotive embedded systems, is an increasing challenge as the
complexity of these systems significantly increases. EAST-ADL is an
architecture description language developed to specify automotive
embedded system architectures at multiple abstraction levels in the
development of safety-critical automotive products. In this paper,
we propose an architecture-based verification technique which
enhances the model-based development process supported by EAST-ADL
by adapting model-checking to EAST-ADL specifications. We employ
UPPAAL as a verification tool to ensure that predicted function
behaviors of the models in EAST-ADL satisfy functional and real-time
requirements. The criteria for this architecture-based verification
is presented and the transformation rules which comply with this
criteria are derived. This enables us to extract the relevant
information from EAST-ADL specifications and to generate analyzable
UPPAAL models. The formal semantics of EAST-ADL is defined which is
essential to automate the verification of EAST-ADL specifications.
Our approach is demonstrated by verifying the safety of the steering
truck system units.
\end{abstract}
%
%

\section{Introduction}
\label{sec:introduction}

EAST-ADL is an architecture description language for the development
of automotive embedded systems \cite{atesst2}. Advanced automotive
functions \cite{svadm07,kg03} are increasingly dependent on software
and electronics. These automotive embedded systems are becoming
progressively complex and critical for the entire vehicle. Among
such factors which effect the complexity are an increasing number of
components and interactions between them as well as
inter-disciplinary communication. Model-based development (MBD) --
i.e., the use of computerized models \cite{chen09} supports
activities or tools performing activities such as architecture
specification, analysis, V\&V, testing and synthesis -- is a means
to manage this complexity and develop embedded systems in a way that
increases safety and quality.

The EAST-ADL modeling approach addresses this issue and provides the
means to integrate engineering information from documents,
spreadsheets and legacy tools into one systematic structure, an
EAST-ADL system model. Its methodology \cite{eastadl-methodology} is
an emerging MBD solution for automotive embedded systems
development, providing support for architecture specification and
constraints at multiple abstraction levels. EAST-ADL expresses the
structure and interconnection of the system. System behavior is
defined based on the definition of a set of elementary functional
entities and their triggers and interfaces. However, the behavioral
semantics inside each elementary functional entity is not specified,
which limits the automatic translation from EAST-ADL models to other
formal models for efficient verification. Instead, the execution of
each function is described with external informal notations
\cite{bannex} and a legacy tool such as Simulink \cite{simulink},
which is deficient in the analysis of timing requirements with
high-level assurance.

In order to alleviate the aforementioned restriction, we aim to
bring in modeling formalisms to the behavioral definitions of
EAST-ADL that enhances the specification and analysis support for
system models in EAST-ADL, especially early V\&V of behavioral
specifications, taking into account time- and behavioral
constraints. We propose an architecture-based verification
technique, which is based on formal constructs enabling automation
of the verification activities by adopting model checking to an
architectural perspective: We first introduce an EAST-ADL based
verification criteria which describes -- 1. the objects of
verification, i.e., constructs of EAST-ADL selected for evaluation,
and -- 2. the conditions these objects have to meet. Furthermore,
the formal semantics of EAST-ADL is defined and it is critical to
automate verification of EAST-ADL specifications. Secondly, we
present semantics mapping rules which comply with this criteria in
order to transform the selected constructs of EAST-ADL into UPPAAL
Timed Automata (TA) models \cite{Berhmann_2291:2010,alurTA}.
Finally, we demonstrate the applicability of model checking in
automotive applications.

This work is organized as follows. Section 2 introduces technology
and background, EAST-ADL and UPPAAL toolkit as used in our approach.
Section 3 presents our approach and methodology to improve formal
analysis and verification capability of EAST-ADL. In section 4, our
method is demonstrated in verifying the safety of the steering
system units design. In section 5, related work is discussed.
Section 6 concludes the paper and plans future work.

\section{Preliminaries}
\label{sec:background}

\subsection{EAST-ADL}
\label{sec:east-adl}

EAST-ADL (Electronics, Architecture and Software Technology --
Architecture Description Language) is an ADL dedicated to automotive
electronic systems resulting from several European projects
\cite{timmo,atesst2}. The language provides support for
architectural specifications at different abstraction levels.

The highest abstraction level, \emph{Vehicle(Feature) level},
characterizes a vehicle by means of its features and requirements.
At \emph{Analysis level}, functionality is realized based on the
features and requirements. These features and requirements are
refined by the decision of logical design with the definition of
logical abstract functions of features and their interactions, and
requirements. The model at this level is used for the analysis of
control requirements, timing constraints, data consistency between
interfaces, hazard identification, etc. \emph{Design level} contains
concrete functional definitions according to the realized logical
design. In particular, functional definitions of application
software, functional abstraction of hardware and middleware are
presented, as well as hardware architecture being captured and
function-to-hardware allocation being defined. At
\emph{Implementation level}, the component architecture is
represented using the AUTOSAR standard \cite{autosar} or similar
specifications.

EAST-ADL has extensions for environment modeling, requirements
specification, timing, dependability and V $\&$ V at every
abstraction level. This modular approach separates the definition of
functional and non-functional aspects such as behavior and related
constraints for such behaviors as timing, and safety. The behavior
annex \cite{bannex} contains extensions to EAST-ADL for a more
fine-grained specification of key behavior attributes and is aligned
with the language but not yet validated, nor included in the base
specification.

In this paper we refer to this behavioral annex (BA) to construct an
analyzable UPPAAL TA model.  The contents of BA such as timing- and
behavior constraints, are utilized to constrain behavioral
definitions in the UPPAAL TA model. Moreover, those constraints as
well as textual requirements are formalized in linear time logics
and are used as UPPAAL queries, which can be verified over the
generated TA model by UPPAAL model checker. Details will be
explained in Section \ref{sec:metholdolgy}.
%

\subsection{UPPAAL}
\label{sec:uppaal}

UPPAAL \cite{Berhmann_2291:2010} is a model checking tool for
modeling, validation and verification of real-time systems. UPPAAL
is used to model a system as a network of Timed Automata
(TA)\cite{alurTA}  -- this tool allows construction of large models
by composing TA in parallel and permits communications between TAs
using share discrete and clock variables and synchronize
(rendezvous-style) on complementary input and output actions, as
well as broadcast actions -- and to verify requirements of the model
by the UPPAAL query language expressed in a subset of Computational
Tree Logic (CTL) \cite{mcbook}. The property specification language
supports safety, liveness, deadlock, and response properties that
can be solved by reachability analysis.

UPPAAL is comprised of three main parts: an editor, a simulator, and
a verifier for modeling, early fault detection (by examination
without exhaustive checking) and verification (covering exhaustive
dynamic behavior) respectively in a graphical environment. UPPAAL
uses the concept of templates for reusability and prototyping of
system components. Each template can be instantiated multiple times
to represent similar TA with varying parameters.

Additional features offered by UPPAAL in comparison to other
model-checking tools are bounded integer variables, committed and
urgent conditions. Committed locations must be left immediately by
the next transition taken in the system. Urgent locations must be
left without letting time pass, but allows interleaving by other
automata. For details of UPPAAL TA, we refer the reader to
\cite{Berhmann_2291:2010}.

Another feature of UPPAAL is the use of a subset of C-code for its
declarations. This is beneficial in the sense that C is one of the
most commonly used languages for real-time systems. Furthermore, it
is possible to use C syntax for EAST-ADL expressions. L.Feng et al.
\cite{lei10} transformed C syntax in EAST-ADL to PROMELA to enable
model checking in SPIN. This is not required in the case of UPPAAL
if the same subset of C is used for both EAST-ADL and UPPAAL.

\begin{figure}[!tp]
  \centering
   \includegraphics[width=3.4in]{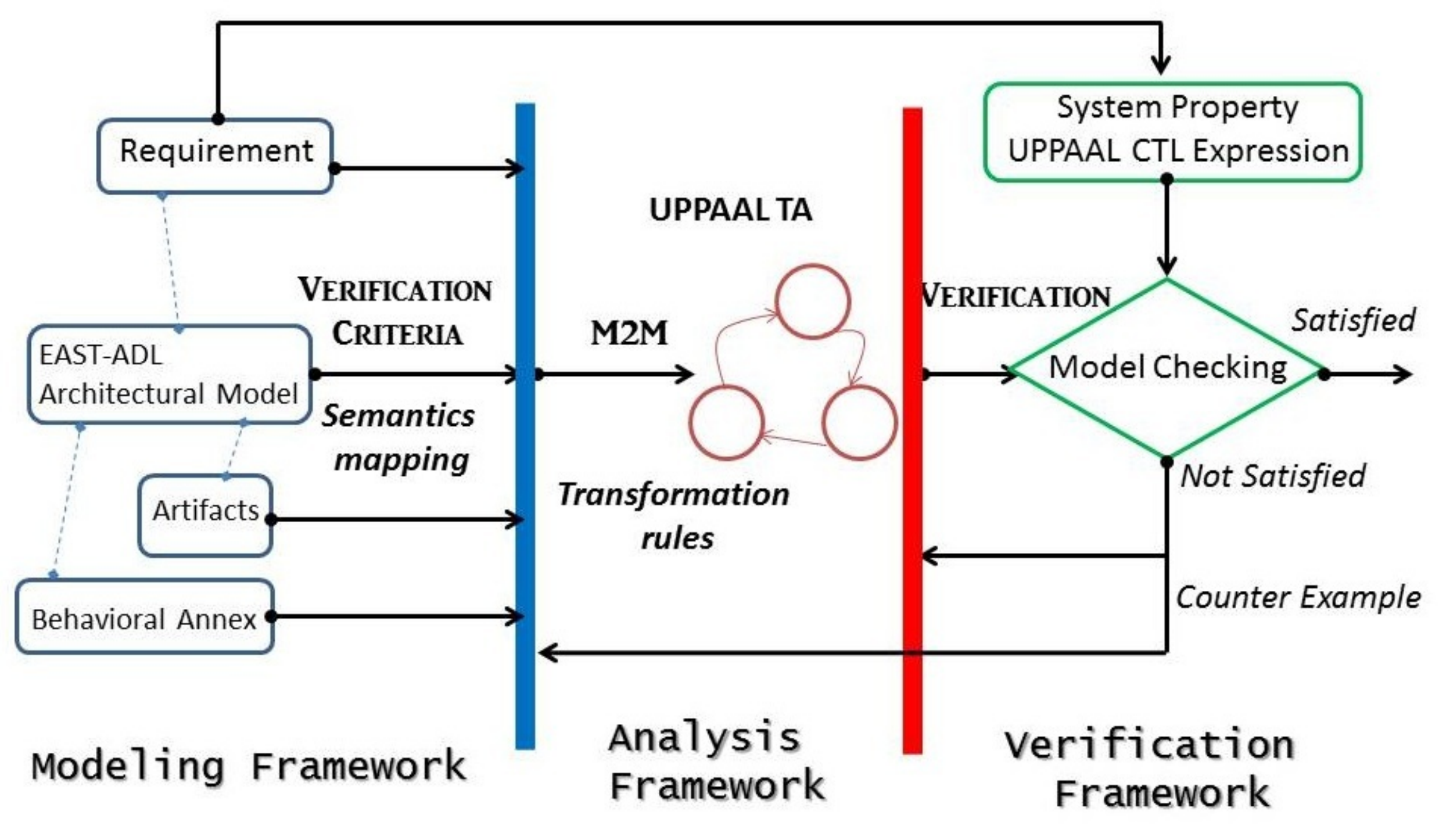}
  \caption{Methodology Roadmap}
  \label{fig:approach}
\end{figure}

\section{Approach and Methodology}
\label{sec:metholdolgy}

To achieve our aforementioned goal in section
\ref{sec:introduction}, we propose a formal approach which
facilitates the verification of system behaviors in EAST-ADL by
using UPPAAL model-checker independently of any hardware constraints
and topology mapping. It focuses mainly on the higher level of
functional behavior of applications at \emph{Analysis level} in
terms of its \emph{Feature level} with three distinct phases --
verification criteria, M2M transformation, and verification (model
checking). We will discuss those phases in more detail in following
sections.

\subsection{Verification Criteria} 
\label{sec:criteria-mapping}

An EAST-ADL-based verification criteria is introduced and applied to
the step from \emph{Modeling Framework} to \emph{Analysis Framework}
in our methodology roadmap Fig.\ref{fig:approach}. Indeed, the
formal semantics of EAST-ADL is defined and it is essential to
facilitate the automation of V\&V technique.

The verification criteria describes the objects of verification,
i.e., constructs of EAST-ADL specifications selected for evaluation,
and the conditions these objects have to meet. The possible objects
are restricted to four different types of constructs:
\emph{Functions}, \emph{Ports} and \emph{Connectors},
\emph{Artifacts}, and \emph{Behavioral Annex}.\vspace{6pt}

\begin{figure}[!tp]
  \centering
   \includegraphics[width=3.4in]{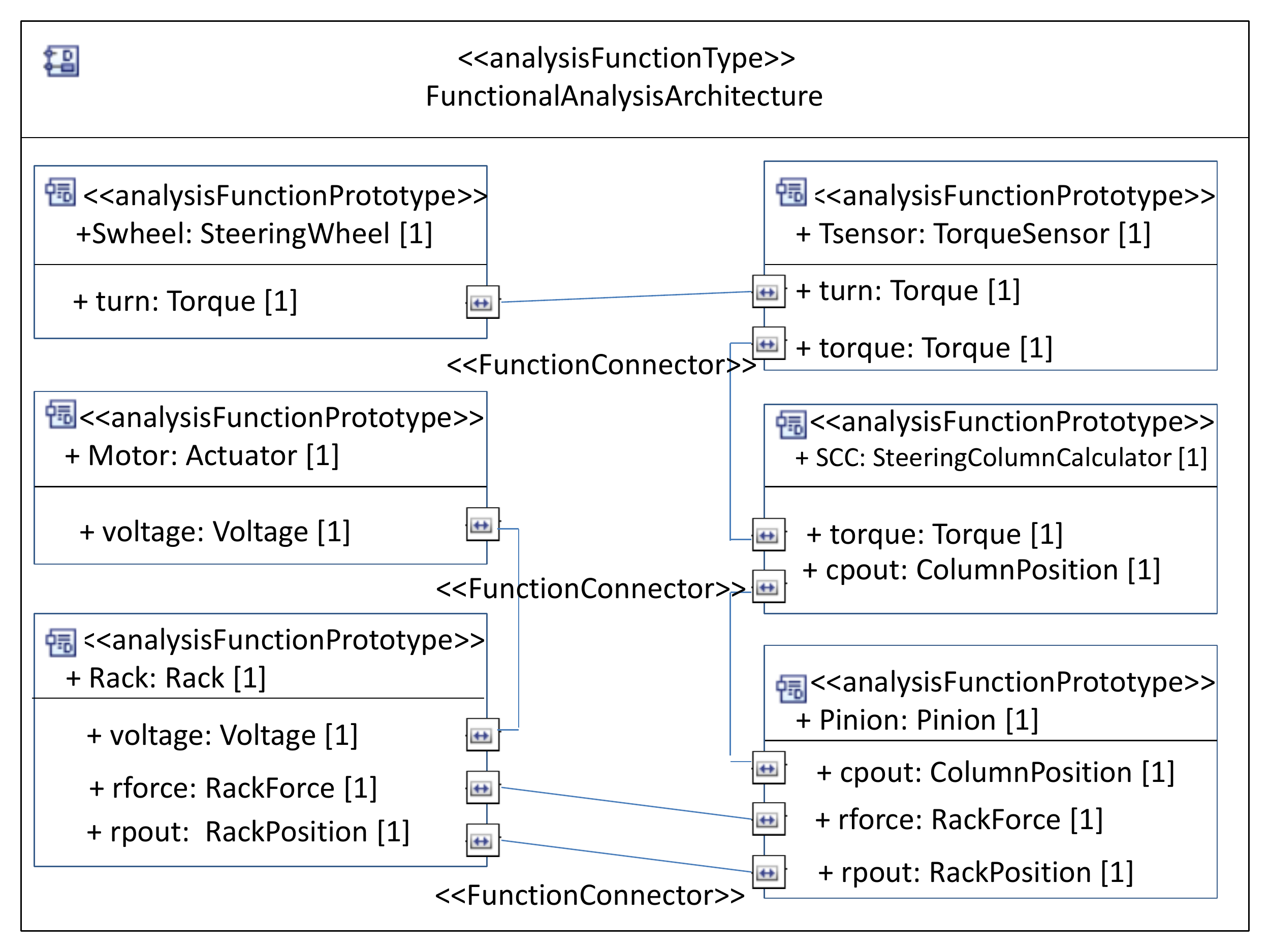}
  \caption{Steering Truck System modeled in EAST-ADL}
  \label{fig:structure}
\end{figure}

\noindent \textbf{Functions:} EAST-ADL refers to components as
\emph{features, functions} or \emph{components}, depending on which
conceptual abstraction level is considered. As we focus on
\emph{Analysis level}, \emph{AnalysisFunctions} (AF) are selected as
abstraction of computational units, i.e. SteeringColumnCalculator,
TorqueSensor, etc. They are illustrated as function entities in
Fig.\ref{fig:structure}. Each AF has its own logical execution,
which is specified in UPPAAL TA and used for verification in this
paper.\vspace{6pt}

\noindent \textbf{Ports and connectors:} There are two types of
function interaction either a \emph{FunctionFlowPort} interaction
whereby a function performs a computation on provided data, or a
\emph{ClientServerPort} interaction whereby the execution of a
service is called upon by another function. The
\emph{FunctionFlowPort} are directional interaction points for
exchange of data between AFs and specified by associated data-types,
and they are connected to other ports via connectors. Both
interactions are validated by checking if the execution of TA and
its synchronization with other TAs are not blocked. Since EAST-ADL
connectors are dependent on the ports of AFs they connect, the
modeling and verifying connectors should describe which AF ports are
connected together, and ensure that data- and control flows
interactions between AFs through the ports are correct
respectively.\vspace{6pt}

\begin{figure}[!tp]
  \centering
  \includegraphics[width=3.3in]{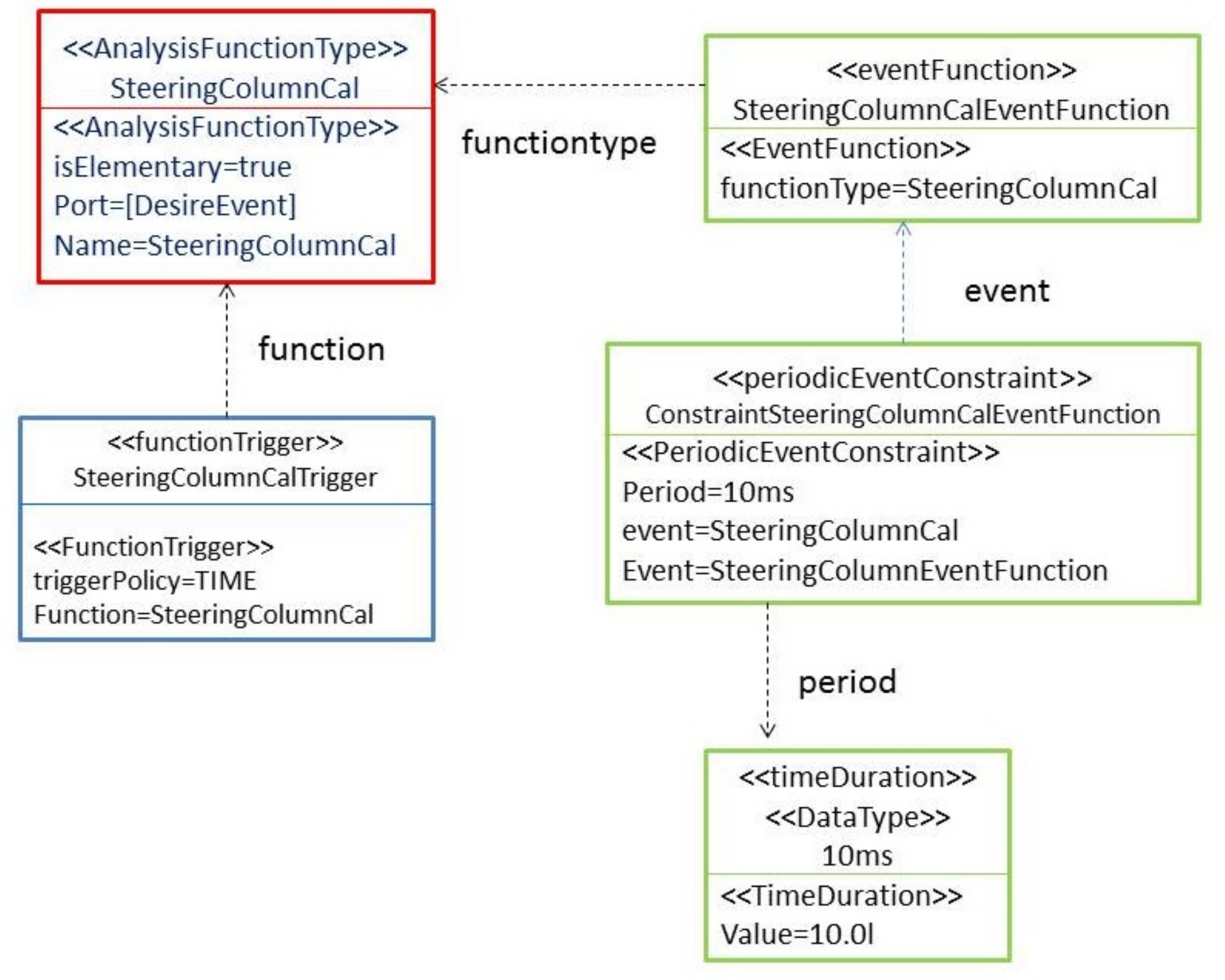}
  \caption{Steering Column Calculator artifacts}
  \label{fig:eastadlspec}
\end{figure}

\noindent\textbf{Artifacts:} Each AF consists of three different
kinds of artifacts: structure, behavior and timing packages. They
are shown as red, blue, and green respectively in
Fig.\ref{fig:eastadlspec} where the \emph{FunctionTrigger} defines
\emph{TriggerPolicy} as either time or event-behavior to specify
timing constraints. According to the \emph{TriggerPolicy}, the AF is
invoked either by time-triggered in which time alone causes
execution to start, or event-triggered, which is caused by data
arrival or calls on the input ports. For each AF, its artifacts are
used to construct corresponding TA and declare how the execution of
TA is dispatched.\vspace{6pt}

\begin{figure*}[tp]
  \centering
   \includegraphics[width=5in]{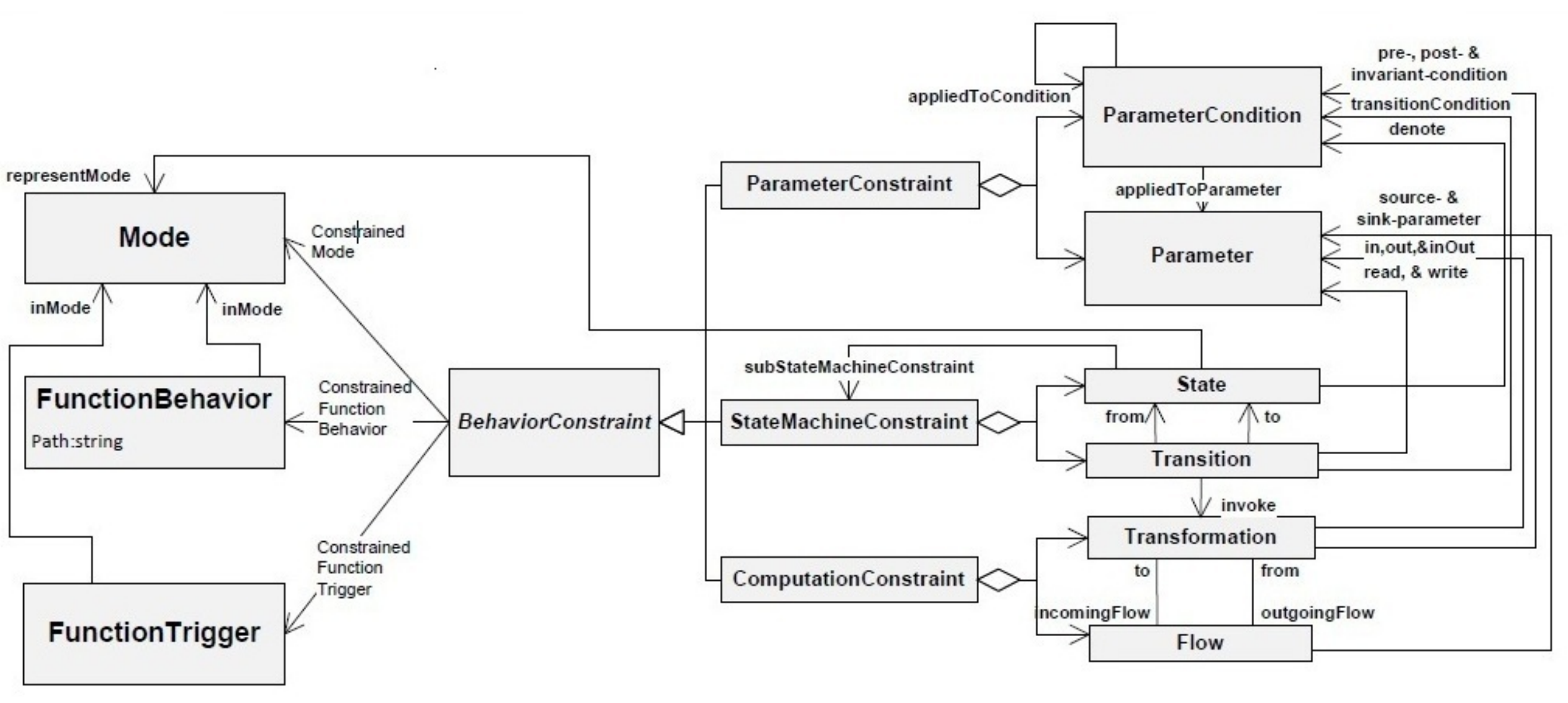}
  \caption{Schematic View of Behavioral Constraints Annex}
  \label{fig:BA}
\end{figure*}

\noindent \textbf{Behavioral Annex (BA):} BA contains behavioral
constraints which are given in three categories: \emph{Parameter
Constraints}, \emph{State Machine Constraints}, and
\emph{Computation Constraints} (See Fig.\ref{fig:BA} with
dependencies). Based on aforementioned artifacts, we further detail
each computation and execution behavior of TA by capturing related
\emph{Parameter-}, \emph{State Machine-}, and \emph{Computation
Constraints} from BA.

In reference to \emph{Parameter Constraints}, we specify -- 1.
quantities of constraints to be processed by the computation and
execution behavior of TA model, and -- 2. conditions under which
each of these quantities is used. The conditions of \emph{Parameter
Constraints} such as pre-, post, and invariant conditions in the
computational modes in EAST-ADL as well as the relations of
parameters (e.g., input and output mapping, event to output mapping)
are used as (i.e. translated into) the corresponding conditions in
TA model. The locations and transitions of TA model are constructed
taking into account \emph{State Machine Constraints}.  The logical
transformation of data in a \emph{FunctionalBehavior} is represented
in TA by referring to \emph{Computation Constraints} or the
conditions of \emph{Parameter Constraints}

According to the objects of verification above, the execution of
\emph{FunctionalBehavior} within an EAST-ADL system can be
considered a combination of multiple end-to-end computation and
execution chains across its AFs under the triggering definitions,
and the behavioral- and timing constraints given in artifacts and BA
respectively. Thus, \emph{evaluating the FunctionalBehavior} in
EAST-ADL is to ensure the two following conditions: \vspace{6pt}
\begin{enumerate}
\item For each AF, its corresponding TA execution such as flow- or
data interaction via synchronization with its environments (other
TAs) is not blocked, i.e., Deadlock-free condition.

\item When the interaction or synchronization is carried out, data
transformation is performed as an expected computation activity on
two sets of parameters. The conditions of those parameters such as
pre-, post-, and invariant conditions must hold before the
transformation starts its execution, after the execution of the
transformation, and must remain unchanged by the execution of the
transformation respectively.
\end{enumerate}
\vspace{6pt}

\noindent To perform a behavioral semantics mapping from the
informal semantics of EAST-ADL architectural model to the formal
semantics of UPPAAL TA model, we formally define behavioral
semantics of EAST-ADL below. This step is critical to automate the
verification technique using UPPAAL model checker. UPPAAL tool
composes each local TA of AF in parallel to a composed TA (network
TA).

Essentially, EAST-ADL model is a tuple $\langle N, CE \rangle$,
where $N$ is a set of \emph{ADLFunctionalPrototypes} AFs, and $CE
\subseteq N \times N$ is the set of connectors between AFs. Output
variables of one AF may be connected to input variables of another
AF. The behavior $\mathcal{B}$ inside an AF, noted ${\llbracket B
\rrbracket}_{AF}$, is modeled as an UPPAAL TA $= \langle L, l_0,
V_C, V_D, E, I \rangle$, where $L$ is a set of locations, $l_0 \in
L$ is the initial location, $V_C$ and $V_D$ is a set of clock and
data variables respectively. $I$ assigns an invariant to each of the
locations. $E$ is a set of edges, represented as $l \move{g,a,u}
l'$, where $l$ is a source location, $l'$ is a destination location,
$g$ is a guard, $a$ is an action, $u$ is an update.

The execution of each AF behavior is determined in terms of
triggering values (conditions) given in BA or artifacts. When the
triggering condition is  \emph{active}, the AF is triggered via its
input trigger port, and its input data ports are mapped to data
variables. $V_D$ in TA are updated with those variables by
\emph{read-input-from-ports} action, noted READ($P_{in}$),
(respectively \emph{write-output-to-ports}, noted WRITE($P_{out}$)),
which are atomic and urgent (in the sense that time is not allowed
to pass when AF reads or writes). AF is initially \emph{idle}  after
the read action it switches to its executing locations until its
internal computation is done. After the write action, which forwards
data in variables via connections from the output ports, the AF
becomes \emph{idle} again and the trigger condition is updated to
\emph{inactive}. The behavior of AF is formally defined as follows:

\begin{definition}[Definition 1]
The behavior of AnalysisFunction AF is a tuple $\mathcal{B} =
\langle L, l_0, l_f, V_D \cup P , V_C, E \cup \{e_r, e_w\} \rangle$
where

\begin{itemize}
\item $l_{0}$ is the initial location (also considered the idle location).

\item  $l_{f}$ is the final executable location such that the edges in $E$
leading out from $l_f$ lead to the $l_0$.

\item $P$ is the set of ports of the component described as $P_{in} \cup P_{out} \cup
P_{trig}$, where $P_{in}$ is a set of input ports, $P_{out}$ is a
set of output ports, $P_{trig} \subseteq P_{in}$ is the set of
trigger input ports.

\item $e_r = l_{0} \move{g,r,u} l'$, if $g$ is triggered,
$r$ is the ''read-input-from-ports'' action, READ($P_{in}$), and $u$
updates $V_D$ with input values $(P_{in} \backslash P_{trig})$.

\item $e_w = l_f \move{g,w,u} l_{0}$, if $g$ is true,
$w$ is the ''write-outputs-on-ports'' action, WRITE($P_{out}$), and
$u$ resets $P_{trig}$ to ''inactive''.
\end{itemize}
\end{definition}

\noindent The TA of a composition $C$, TA($C$), is defined as a
network of local TA. For $AF_i$ and its corresponding component $C_i
\in C$, the write action in TA($C_i$) is extended to update the
input ports (noted $P_{in.j}$) of a target component $C_j \in C$
according to interconnections from the out ports of $C_i$ (noted
$P_{out.i}$). An interconnection connects a source port $p \in
P_{out.i}$ to a target port $p' \in P_{in.j}$ whenever variables in
$P_{in}$ of $C$ are enabled in a way that if $p'$ is a trigger port
then $p'$ is activated, otherwise $p' = p$. The edges $e$ of
TA($C_i$) are explained with extended write actions as follows.

\begin{definition}[Definition 2]
The behavior $\mathcal{B}$ inside $AF_{i}$, $\llbracket B
\rrbracket_{AF_i}$, is TA($C_i$) $=$ $\langle$ $L$, $l_0$, $l_f$,
$V_D$, $V_C,$ $\{XWRITE_{i}(e)$ $\mid$ $e \in E\}, I \rangle$ such
that
\begin{itemize}
\item $XWRITE_i$ ($l \move{g,a,u} l'$) $\triangleq$  ($l \move{g,w,u}
l'$ ; $WRITE$($P_{out.i}$)), if $a=w$ and $g$ is triggered (holds).\\
Note that $;$ is defined as sequential execution
%
\item $XWRITE_i$ ($l \move{g,a,u} l'$) $\triangleq$  $l \move{g,a,u} l'$
, for $a \neq w$
\end{itemize}
The automata TA(C) is then the network of each TA($C_{i}$) for $C_i
\in C$.
\end{definition}

\noindent An environment is modeled as $TA_{Env}$ in a similar way.
The resulting composition is thus defined as the network $TA(C)
\times TA_{Env}$, where any edge in $TA_{Env}$ updating ports
$P_{in}$ of $C$, is extended with an update $WRITE(P_{out.Env})$.
This is similar to the adaption of the $XWRITE$ action that is used
to build $TA(C_{i})$ in Definition 2.

\subsection{M2M Transformation}
\label{sec:transformation-rule}

This model-to-model transformation step, called M2M, is to transform
EAST-ADL specifications to UPPAAL models under the verification
criteria. This step is illustrated as a M2M direction in
\emph{Analysis Framework} in Fig.\ref{fig:approach}. Automated
formal verification is not feasible directly on EAST-ADL
specification since EAST-ADL lacks formal and implemented semantics.
In order to perform automated formal verification, we first formally
mapped the execution semantics of EAST-ADL to UPPAAL TAs
(Ref. Def 1 and 2 in Section \ref{sec:criteria-mapping}). 

In this section, we present transformation rules which helps to
extract executable UPPAAL TAs from EAST-ADL specifications. The
derived UPPAAL models comply with artifact packages, BA and
functional requirements of EAST-ADL model under the verification
criteria. A few assumptions are made in order to simplify the
transformation procedure:

\begin{itemize}
\item We focus on \emph{FunctionalAnalysisArchitecture} (FAA) at the
analysis level of EAST-ADL where each AF is associated with
\emph{AnalysisPorts}. Except for FAA, no function type is allowed to
have instances for other function types.

\item Unlike the original EAST-ADL specification, ports' events are
related not only to triggering but also to synchronization. This
implies that sending and receiving data from a port is also
considered an event although it is not used to trigger a particular
function. This assumption is required for the verification of
end-to-end timing property using reaction time constraints.

\item The communication channels between different processes in UPPAAL
is of broadcast type. This complies with EAST-ADL specifications
where each elementary AF follows run-to-completion execution with
non-blocking semantics and ports are over-writable buffers.
\end{itemize}

\noindent In terms of these assumptions, we first present a general
transformation scheme which can be described as follows.\vspace{6pt}

\noindent \textbf{M2M General Transformation Scheme:} For each
EAST-ADL system (FAA), an UPPAAL system consisting of a set of
processes is created. For each elementary AF, an UPPAAL process is
defined whereby a local clock is declared. This local clock is used
to specify the execution time of AF.

Each connector between AF ports is transformed to an UPPAAL
broadcast channel. Global and local variables of processes are
defined regarding \emph{Parameter Constraints} in BA. The
value-passing action between two AFs is performed using global
variables via broadcasting channels between the corresponding UPPAAL
processes. Control- or data flow behaviors of AFs follow the
computation activities on two sets of quantities in terms of the
\emph{Computation}- and \emph{StateMachine} constraint, i.e.,  pre-,
port-, and invariant parameter conditions in BA.

Secondly, and more importantly, we present three transformation
rules, \emph{Time-Triggering rule}, \emph{Event-Triggering rule},
and \emph{End-to-End reaction time rule}. Each proposed
transformation rule is specified in UPPAAL TA and a schematic
graphical-structure view of the corresponding EAST-ADL is
illustrated below.\vspace{6pt}

\begin{figure}[!htp]
\begin{center}
  \includegraphics[width=3.3in]{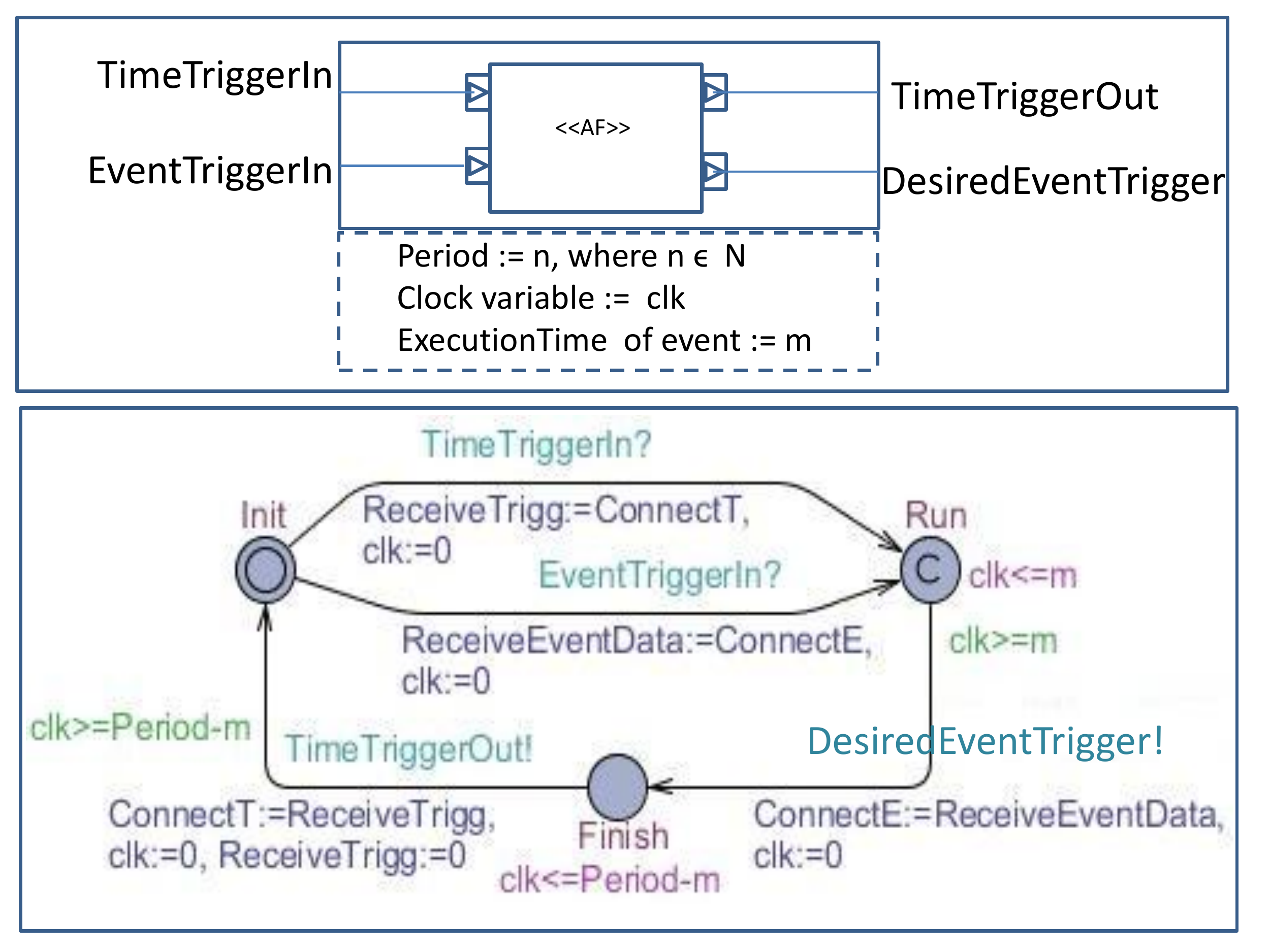}
  \end{center}
  \caption{Time-triggered rule}
  \label{fig:ttp}
\end{figure}

\begin{figure}[!hpt]
\begin{center}
  \includegraphics[width=3.3in]{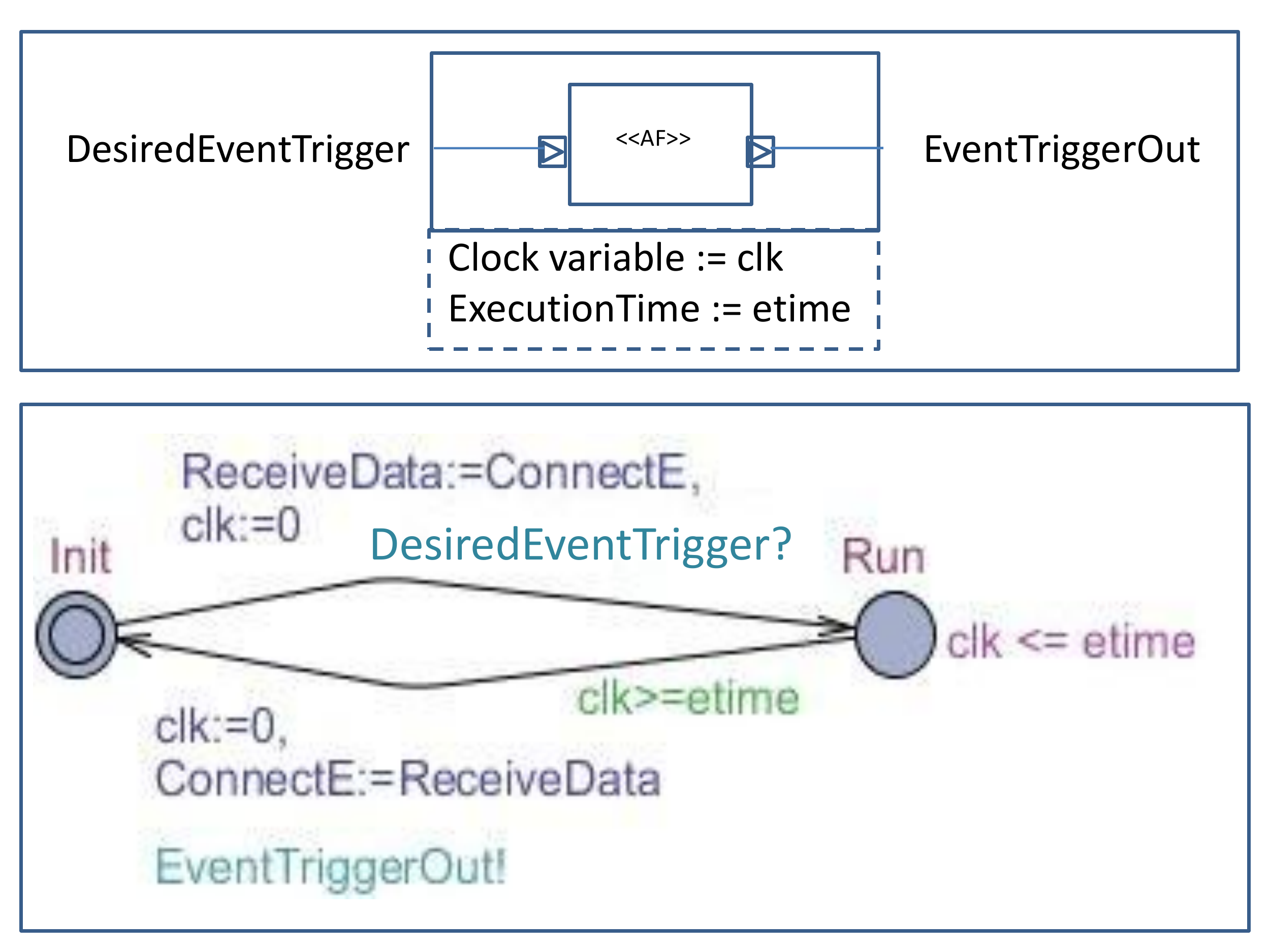}
  \end{center}
  \caption{Event-triggered rule}
  \label{fig:etp}
\end{figure}

\noindent \textbf{Time-Triggering rule:} This rule transforms the
periodic behaviors of AF.  From the given artifacts and BA of
EAST-ADL model, if related FunctionTriggerPolicy is TIME and its
TriggerPeriod defines the periodic triggering time $(Period := n)$ ,
i.e., this AF is invoked every $n$ time, then the EAST-ADL model
which is illustrated as the upper view in Fig.\ref{fig:ttp} is
transformed to the corresponding TA which is depicted as the lower
view in Fig.\ref{fig:etp}.

An event-triggered function may be invoked during the
time-triggering period due to data arrival or call from or to other
AFs. If this happens, then  an invariant $clk$ $\leq$ $m$ ($m$ is
$ExecutionTime$ of the event) is added to $Run$ location. An
invariant $clk \leq Period - m$ in \emph{Finish} location specifies
the expiry of the fixed-periodic triggering time after the
$ExecutionTime$ of event-triggered functions. For each transition,
the local clock is reset. For each port associated with its
connector, a related broadcast channel is defined as the port name
$TimeTriggerIn?$ and $TimeTriggerOut!$ representing time-triggering
input- and output ports or $EventTriggerIn?$ and
$DesiredEventTrigger!$ representing event-triggering input- and
output ports respectively.

A global integer variable $ConnectT$ and a local integer variable
$ReceiveTrigg$ are declared. $ConnectT$ is assigned the value of 1
initially. When a time-triggering signal is received through
$TimeTriggerIn?$, the local variable $ReceiveTrigg$ is assigned the
value of $ConnectT$. When the signal is sent out through
$TimeTriggerOut!$, the global variable $ConnectT$ is assigned the
value of $ReceiveTrigg$, and then the local variable $ReceiveTrigg$
is reset to 0.

Similarly, the value of $ReceiveEventData$ is passed from the
current TA in Fig.\ref{fig:ttp} to the variable $ReceiveData$ of TA
in Fig.\ref{fig:etp} using the global variable $ConnectE$ through
the broadcast channel $DesiredEventTrigger$. Location $Run$ in
Fig.\ref{fig:ttp} is committed to ensure that no other TAs can
assign their local variables with the value of $ReceiveEventData$
before the TA in Fig.\ref{fig:etp} sets $ReceiveData := ConnectE$.
Once the periodic triggering time is expired, i.e., $clk$ $\geq$
$Period - m$, then
 $Time\_TriggerOut!$ is broadcast. Therefore, this rule can
facilitate the synchronization of various clocks in a design and
avoid unpredictable delays.\vspace{6pt}

\noindent \textbf{Event-Triggering rule:} From the given artifacts
and BA of EAST-ADL model, for any AF if related
FunctionTriggerPolicy is EVENT and its TriggerPeriod defines the
execution time of the invoked event $(ExecutionTime := etime)$, then
the EAST-ADL model which is illustrated as the upper view in
Fig.\ref{fig:etp} is transformed to the corresponding TA, which is
depicted as the lower view.  The defined execution time ($etime$) is
added as an invariant to $Run$ location to specify the expiry of the
$ExecutionTime$ of event-triggering function. For each transition,
the local clock is reset. For each port, a relevant broadcast
channel is defined as the port name $DdesiredEventTrigger?$ and
$EventTriggerOut!$ representing event-triggering input- and output
ports respectively. Value passing from current TA to the others is
represented similar to the aforementioned one in the Time-triggering
rule above.

The executable event flows from $Run$ location to $Init$ location
can be refined into corresponding behavioral automaton where $Run$
and $Init$ locations being the initial location and the final
location respectively. The refinement is made straightforward by
adding more $Running$ locations, transitions, and associated
variables as well as broadcasting channels between $Run$ and $Init$
locations. They represent more interactions between the current AF
and other AFs. The timing properties of the basic and refined model
should be identical, because the newly introduced locations
represent the former original location, and the union of the new
locations clock invariants (union of individual execution time)
should fulfill the former invariant execution time ($clk \leq
etime$). \vspace{6pt}

\noindent \textbf{End-To-End reaction time rule:} This rule
transforms the End-To-End reaction time behavior between two AFs
into TAs. The reaction time constraint which is given in BA is first
represented as the \emph{bounded response time property} formula of
the form $AG(f1$ $
\Rightarrow_{\leq T}$ $f2)$, meaning that \emph{if a
request (f1) becomes true at a certain time point, a response (f2)
must be guaranteed to be true within a time bound (T)}.

\begin{figure}[!htp]
\begin{center}
  \includegraphics[width=3.4in]{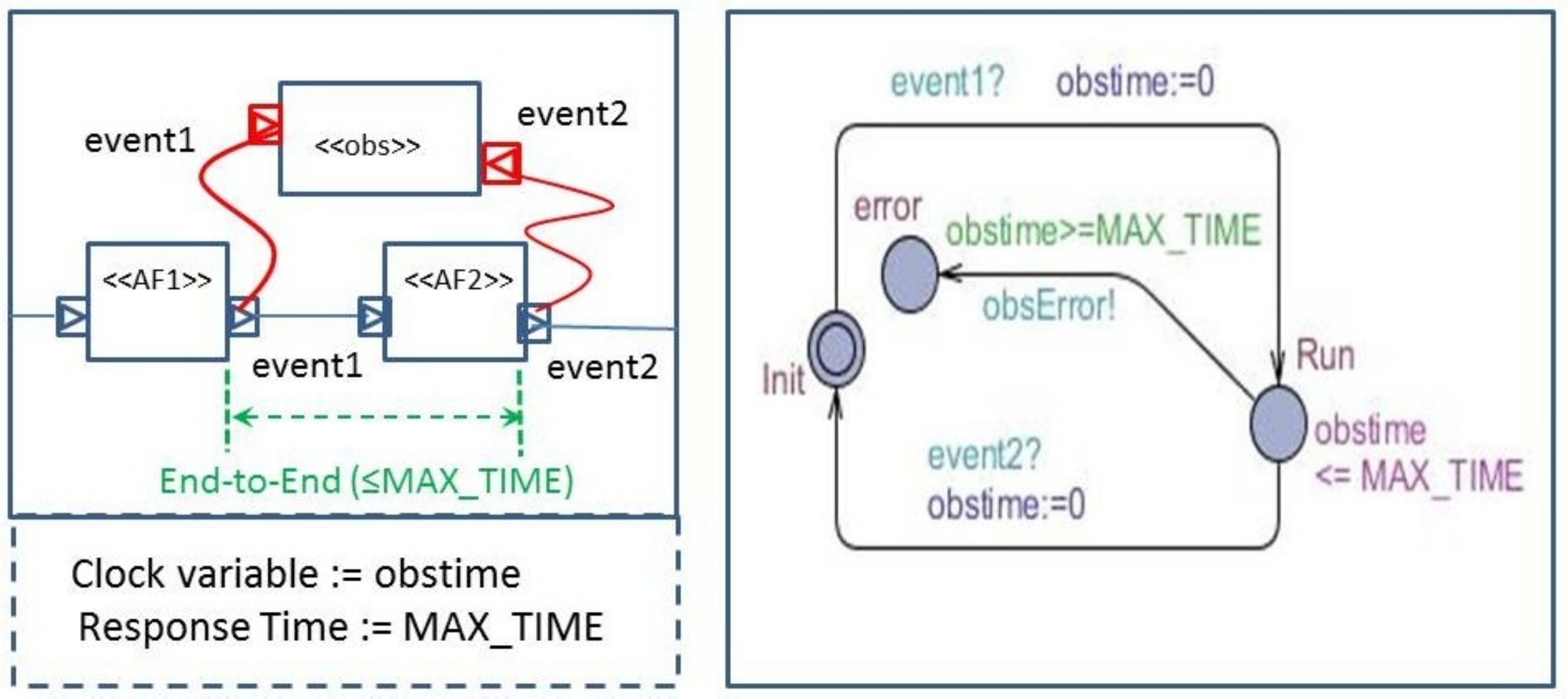}
  \end{center}
  \caption{End-To-End reaction time rule}
  \label{fig:ete}
\end{figure}

In order to verify this time constraint property, we take a similar
approach from our early experiment in \cite{ksafecomp11}, which
shows how to check bounded liveness properties with a certain
syntactical manipulation on the system model. An observer TA (see
righthand view in Fig.\ref{fig:ete}), which restricts the bounded
response time ($MAX\_TIME$) is constructed. This observer TA
contains a clock constraint ($obstime \leq MAX\_TIME$) and an
\emph{error} location, which violates the bounded response time
condition.

One synchronization channel $event1?$ is added to the transition
from $Init$ to $Run$ in order to deal with a request event of AF1.
Another synchronization channel $event2?$ is added to the transition
from $Run$ to $Init$ in order to handle a response event of AF2.
Afterwards, the observer is syntactically composed with the actual
AFs (TAs), say AF1 (TA1) and AF2 (TA2), then we checked if its error
location can never be reached from any location of AF1 (TA1) and AF2
(TA2).

\subsection{Verification: Model Checking}
\label{sec:verification}

The execution of each \emph{FunctionBehavior} in EAST-ADL is
transformed to UPPAAL TA which holds the verification criteria and
its composition is considered as the network TA. The formal
semantics of EAST-ADL has been investigated, where UPPAAL TA is
used. The entire system (network TA) is considered in terms of a
timed transition system, then this entire system is verified by
UPPAAL model-checker. Quality requirements (e.g, timing, safety,
deadlock freedom) in terms of functional requirements (e.g,
behavioral constraints, timing constraints), see Fig.1-Requirement
aspect, are formalized in linear time logics based on the UPPAAL
logic, which can be verified over the transformed TA model by UPPAAL
model checker. This step is shown as \emph{Verification Framework}
in Fig.1.

In particular, the quality requirements are derived from a given
system's textual descriptions. One may verify certain delay,
reaction and synchronization constraints (i.e, overall behavioral
constraints of a system) according to the quality requirements. For
example, a plausible reaction constraint is 250 ms. In contrast,
functional requirements describe particular constraints of a
function such as timing constraints and trigger elements linked to
an AF entity, i.e., triggering policy and parameter constraints in
artifacts and BA.

UPPAAL model checker verifies those two types of requirements as
safety properties in a way that (1). if a property is satisfied by
the target model, then a functional requirement linked to an AF is
updated to a satisfy relation and generic constraints of the AF are
stored as valid invariants in the V$\&$V structure (VVOutcome linked
to the explained requirement, VVCase, etc) of the EAST-ADL model.
(2). If a property is violated (depicted as \emph{Not Satisfied}
arrows in Fig.1) then UPPAAL model checker returns some
counterexamples that can help analysts to refine the behavioral
constraints of the system model or modify generic constraints in BA,
and identify correct constraints for the AF that it concerns. Thus,
the models in EAST-ADL are updated with the timing assumptions
analysts make as well as the analysis results.

\section{Experiment} \label{sec:experiment}

Our approach applied and demonstrated on a case study, the Steering
System Units (SSU), from our industrial partner VOLVO. It is first
modeled using Papyrus UML \cite{papyrus} for EAST-ADL in the ATESST2
project \cite{atesst2}. The SSU Papyrus UML model consists of six
AFs and the interfaces of the AFs are expressed in terms of ports
and connections as depicted in Fig.\ref{fig:structure}. Each of
these AFs can be further decomposed into smaller or more detail
\emph{DesignFunctions} or \emph{Components} depending on the level
of abstraction in EAST-ADL. The functionalities (AFs) of SSU used
within the current work are as follows: \vspace{0.08in}

\begin{itemize}
\item \textbf{Steering Wheel} is manipulated by the driver and reads
such driver input;
\item \textbf{Torque Sensor} senses the steering wheel position and
sends a desired torque to the steering column calculator;
\item \textbf{Steering Column Calculator} calculates the torque
and positions required for the pinion and sends the calculated
values to the pinion;
\item \textbf{Pinion} senses a rotational motion in terms of the
calculated values from the steering column calculator and sends a
desired torque force as well as a desired rack command to the rack;
\item \textbf{Rack} converts a rotational motion into a linear
motion based on the received values and commands from the pinion;
\item \textbf{Actuator} performs the actual steering of each wheel.
\end{itemize}
\vspace{0.08in}

\begin{figure}[!pt]
\begin{center}
  \includegraphics[width=3.2in]{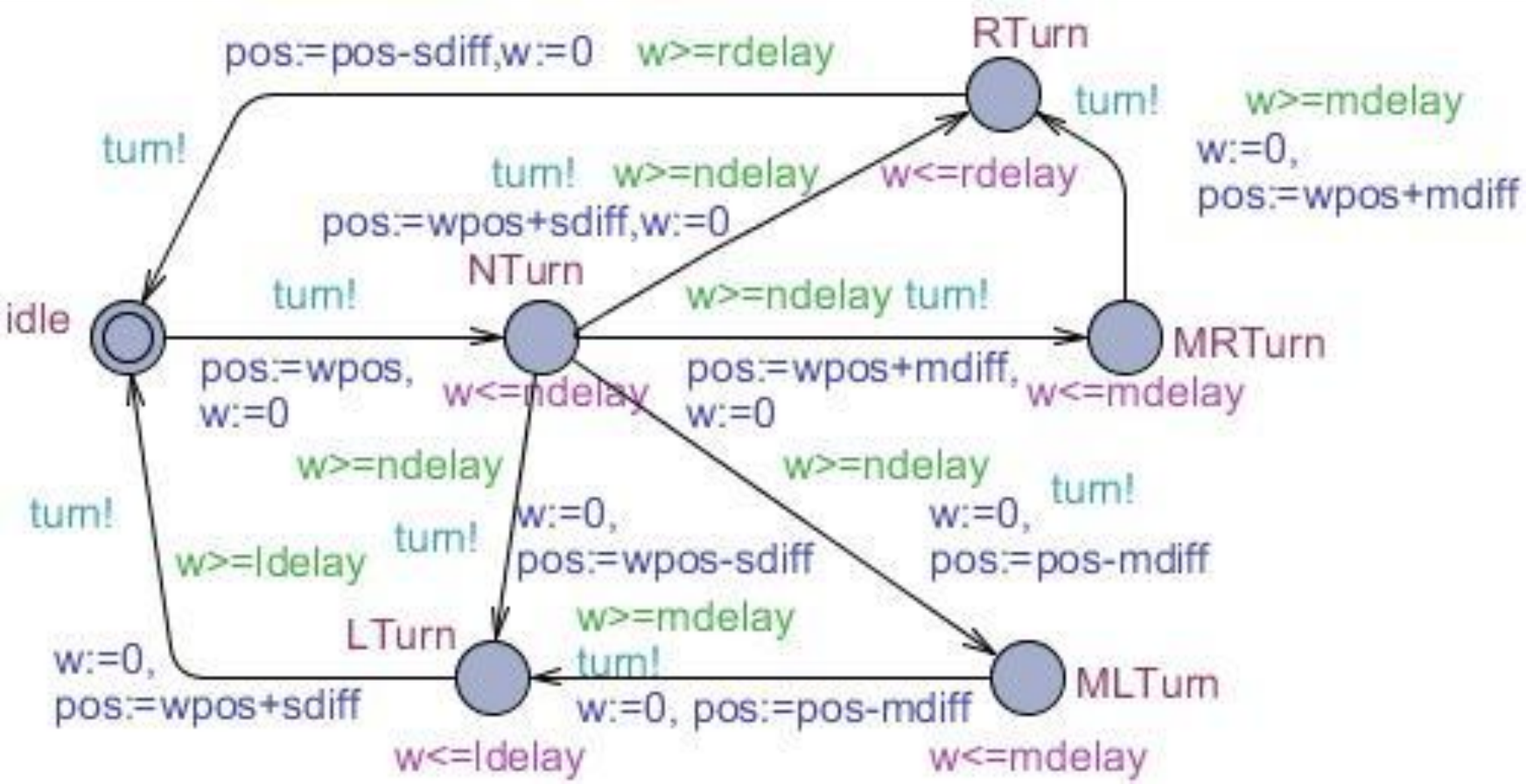}
  \end{center}
  \caption{Steering Wheel UPPAAL TA}
  \label{fig:swTA}
\end{figure}

\begin{figure}[!pt]
\begin{center}
  \includegraphics[width=2.3in]{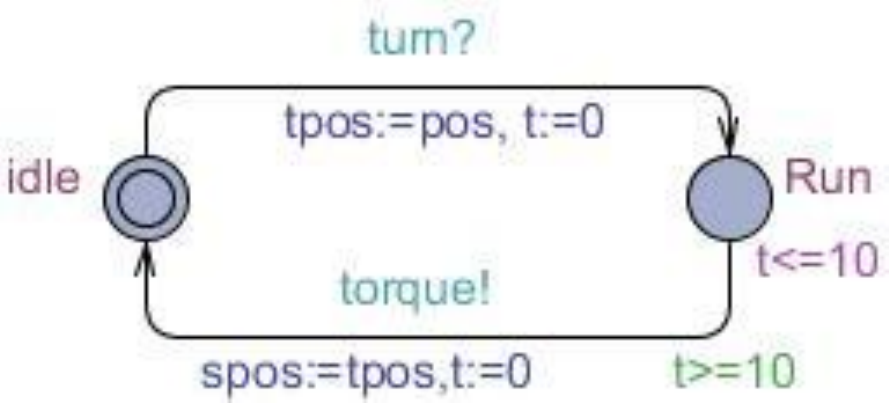}
  \end{center}
  \caption{Torque Sensor UPPAAL TA}
  \label{fig:tsTA}
\end{figure}

\begin{figure}[!pt]
\begin{center}
  \includegraphics[width=3.2in]{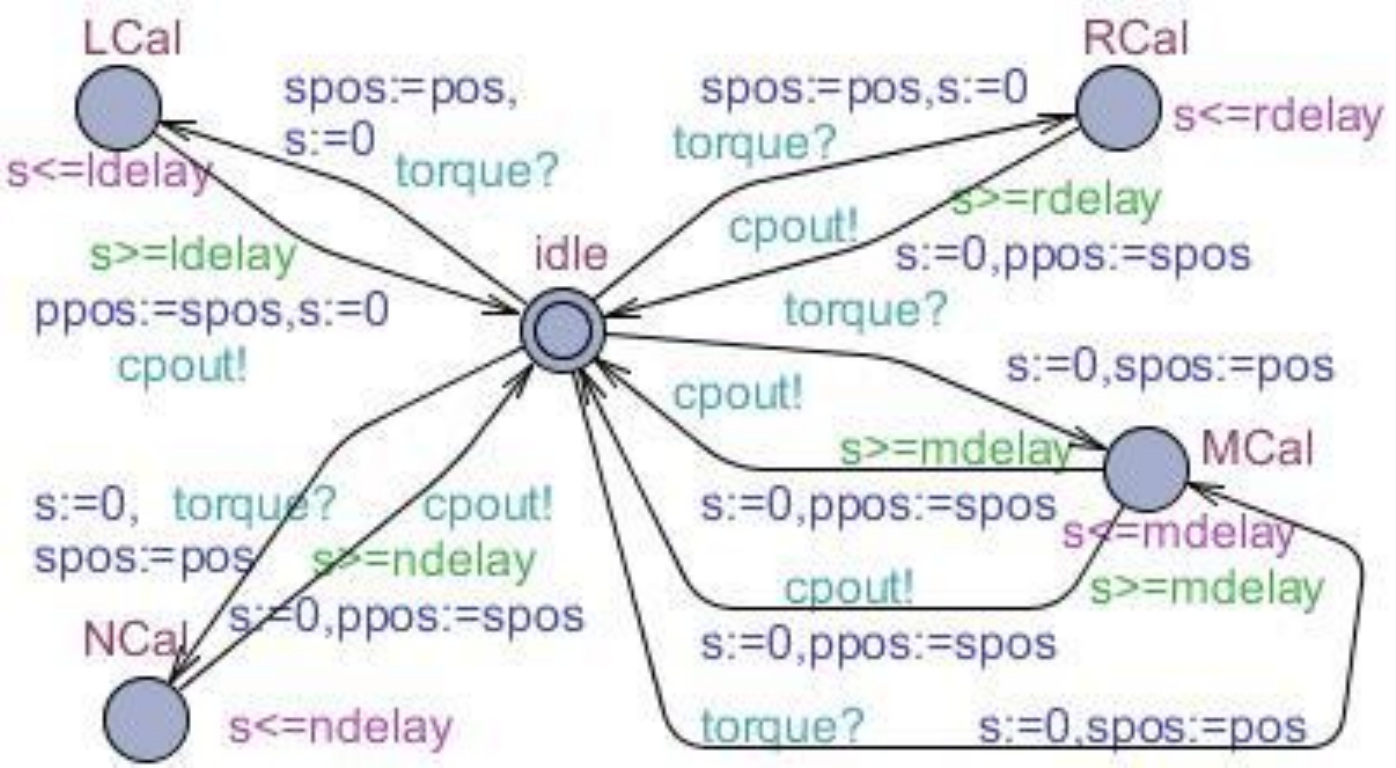}
  \end{center}
  \caption{Steering Column Control UPPAAL TA}
  \label{fig:sccTA}
\end{figure}

\noindent The SSU Papyrus UML model at \emph{Analysis level} in
EAST-ADL is transformed to an UPPAAL model under the verification
criteria: For AFs, \emph{FunctionBehaviors} are semantically
translated into a network of TAs in terms of their corresponding
artifacts and BA. The functional and quality requirements of the
system were given as either informal description in ATESST2 project
case study reports or as timing- or behavioral constraints
requirement entities linked to AFs from BA. Due to space limitation,
here we present only the UPPAAL TAs of \emph{Steering Wheel},
\emph{Torque Sensor} and \emph{Steering Column Calculator} shown in
Fig.\ref{fig:swTA}, Fig.\ref{fig:tsTA} and Fig.\ref{fig:sccTA}
respectively.

The requirements formalized in UPPAAL logics over the network of TAs
are verified by model-checking with some assumptions regarding
timing: there is a data flow from a steering wheel to a motor
actuator. The functions are periodic and mutually unsynchronized. A
perfect clock is assumed in the sense that it generates periodic
triggering in order to activate (run) the AFs with a periodicity of
one time unit. Each function has its execution time which is modeled
with an execution delay location in its TA. Based on those
assumptions, properties of safety, deadlock freedom and liveness are
verified successfully.

We verified 25 properties of the system. The evaluation conditions
mentioned in Section \ref{sec:metholdolgy} is referenced to
formalize the properties. A list of selected properties is given
below and their verification results are established as
valid:\vspace{6pt}

$\backslash \ast$ $\ Definition$ $of$ $each$ $AnalysisFunction$

$\ \ C1 = Steering$ $Wheel$

$\ \ C2 = Torque$ $Sensor$

$\ \ C3 = Steering$ $Column$ $calculator (SCC)$

$\ \ C4 = Pinion$

$\ \ C5 = Rack$

$\ \ C6 = Actuator$ $\ast \backslash$\vspace{6pt}

\begin{itemize}
\item \emph{Deadlock freedom}: $A[\ ]$ \footnote[1]{A[\ ] $P$: ''$P$ holds for
any reachable configuration'' is written $A[\ ]$ in UPPAAL format}
not deadlock

\item \emph{Leads-to property} based on the internal variables of
AFs: Every time Steering Column Calculator (SCC) is invoked by
Steering Wheel or Torque Sensor, it will eventually calculate column
positions for Pinion according to the steering position and torque
force.

    \begin{itemize}[]
    \item $C3.Idle$ $\longrightarrow$
     $(C3.RCal$ $\lor$ $C3.LCal$ $\lor$ $C3.NCal$ $\lor$ $C3.MCal)$
    \end{itemize}

\item \emph{Leads-to property} based on the interactive values via ports:
If Steering Wheel sends out its position value then the value should
be received by SCC.

    \begin{itemize}[]
    \item $(C1.RTurn$ $\land$ $C1.wp == 1)$ $\longrightarrow$
    $(C3.RCal$ $\land$ $C3.cp == 1)$
    \end{itemize}

\item \emph{State correspondence check}: One internal location of an AF
corresponds to what is happening in the locations of other
environment AFs. While Steering Wheel is turning right, neither the
left-column-calculating mode of SCC nor the left-rotation mode of
Pinion is executing.

    \begin{itemize}[]
    \item $A[\ ]$ $C1.RTurn$ $\Rightarrow$ $(\neg$ $C3.LCal$
      $\land$ $\neg$ $C4.LRot)$
    \end{itemize}

\item \emph{Execution time property}: Each AF should execute
within its given local execution time, $0 \leq clk \leq 2$.

    \begin{itemize}[]
    \item $A[\ ]$ $C6.Run$ $\Rightarrow$ $(0$ $\leq$ $C6.clk$ $\leq 2)$
    \end{itemize}

\item \emph{Bounded liveness property}: When Steering Wheel is activated,
Actuator reacts timely under its given time bound (MAX$\_$TIME) as a
failsafe against serious accident.

    \begin{itemize}[]
        \item $A[\ ]$ $C1.RTurn$ $\Rightarrow$ ($\neg$ $ObsTA.error$ $\land$
        $C6.Run)$
    \end{itemize}

In other words, if Steering Wheel is invoked, it should not reach
the error location of the observer TA, which violates the
MAX$\_$TIME bounded condition, while Actuator is executing.
\end{itemize}

\noindent Search order is breadth first and uses conservative space
optimization. The state space representation uses difference bound
matrices (DBM). Verifying properties takes an average of around 2
seconds per verified property on an Intel T9600 2.80 GHz processor.
The UPPAAL model checker needs to explore a maximum of 38166 states.
The maximum is reached when verifying properties where the model
checker needs to explore the whole sate space (such as the
non-existence of a deadlock situation).

\section{Related Work}
\label{sec:related-work}

For safety-driven system development in the automotive domain,
feature- and architecture based analysis is prescribed by ISO
standard as the state-of-the-art approach to functional safety.
However, at an early stage it is difficult to see function
dependencies that would result in updated function requirements.
Therefore,  A. Sandberg et al. \cite{henrik10} provide one approach
that performs iterative analysis to manage changes in the safety
architecture at analysis level and still meet function specific
safety goals derived at vehicle level. In comparison to our work,
their main concern is to define the semantics for requirement
selection in order to ensure correct inclusion of requirements for a
function definition. There is no formal modeling and verification
approach to the behavioral definition of EAST-ADL.

L. Feng et al. \cite{lei10} propose a reference behavior modeling
approach which employs UML activity diagram, C code and SPIN to
study the execution logic of the components and their communication
protocol in EAST-ADL.  Thus the requirements on the system design
can be verified by model checking. In contrast to our work, there is
no notion for the timing constraints in the behavior model. Indeed,
formal analysis on the real-time properties of the behavior model is
not considered at all.

The TIMMO project \cite{timmo} introduces TADL (Timing Augmented
Description Language). The language and its methodology
\cite{timmo-methodology} is aligned with EAST-ADL and AUTOSAR
\cite{autosar} for timing analysis aspects. In contrast to the TIMMO
effort focused on its timing extension, our work focuses on the
behavioral aspects complying with the timing extension. Therefore,
our work contributes towards the development of the EAST-ADL
behavioral extension.

Our earlier work \cite{ksafecomp11} verifies EAST-ADL models using
UPPAAL-PORT\footnote[1]{http://www.uppaal.org/port}. The work uses
mainly structural information of EAST-ADL and its requirement, where
models are manually translated into SAVE-CCM \cite{saveccm}, which
is the required input format for UPPAAL-PORT. In contrast to that
work \cite{ksafecomp11}, we consider recent additions to EAST-ADL,
including both the timing and behavior extensions by analyzing
application internal behaviors specified in behavioral annex as well
as using artifact packages. Furthermore, we eliminate the need for
intermediate formalisms like SAVE-CCM and target direct
transformation of EAST-ADL specifications to UPPAAL TAs.

Another widely used ADL within both industry and the research
community for architectural modeling and analysis of time-critical
software intensive embedded systems is Architecture Analysis and
Design Language (AADL) \cite{aadl}. S. Bj\"{o}rnander et al.
\cite{aadl-verification} propose an approach to formal and
implemented semantics of AADL, where the Timed Abstract State
Machine (TASM) \cite{tasm} language is used as the formal
underpinning.  We take a similar approach by transforming EAST-ADL
constructs to Timed Automata, especially UPPAAL TA, in order to
allow tool-supported automated simulation and verification of
EAST-ADL specification by using UPPAAL model checker. The result of
our work contribute towards the development of EAST-ADL behavior
extension \cite{bannex} and further refinement of the existing
behavior extension.

\section{Conclusion and Future Work}
\label{sec:conclusion}

In this paper, we have studied the use of formal modeling and
verification techniques at an early stage in the development of
safety-critical automotive products which are originally described
in the EAST-ADL architectural language. We have proposed an
architecture-based verification technique which brings formal
modeling formalisms to the behavioral definitions of EAST-ADL to
capture the execution flows inside each functional entity and their
complex interactions and verify them by using the UPPAAL model
checker.

This architecture-based modeling, analysis and verification
methodology is suitable for a higher abstraction level, the current
work focuses on the EAST-ADL analysis level, but can be extended to
the design level. This will require only a change the type of
functions and prototypes in the modeling process.

This paper has furthermore presented the criteria of EAST-ADL based
verification technique. This criteria describes -- 1. the selected
constructs of EAST-ADL such as artifacts, structural Papyrus models,
and BA as well as requirements which are utilized for the purpose of
generating UPPAAL TA models or UPPAAL queries, and -- 2. the
conditions these objects have to meet.

Moreover, the formal syntax and semantics of EAST-ADL are defined,
and this is essential for automated verification of EAST-ADL
specifications, i.e., the formal definition enables the EAST-ADL
model to be automatically transformed to models of other established
tools for V\&V support.  Indeed semantics mapping rules are
presented between EAST-ADL specifications and UPPAAL TAs under the
verification criteria. These rules enable us to generate analyzable
UPPAAL models from the EAST-ADL specifications.

These contributions improve behavior modeling, verification and
analysis capability of EAST-ADL. To demonstrate the applicability of
our approach, the steering system units case study has been
translated from the EAST-ADL Papyrus model into the UPPAAL model. By
employing the UPPAAL model checker, the verification of expected
safety and liveness properties for this system are fully automated.

In future works, we plan to -- 1. include more elaborate
verification of non-functional properties, and more refined
configurations of the generated model. For example, minimizing the
use of certain resources, such as CPU, energy, memory, etc, while
preserving functional correctness, timing requirements and other
resource constraints. The results presented here are promising steps
towards these goals, -- 2. investigate schedulability analysis of
timed-component models (source models from EAST-ADL) using UPPAAL in
order to guarantee that the applied scheduling principle ensures
that the timing deadlines are met. The planned enhancement includes
consideration of functional design prototypes of EAST-ADL models in
Papyrus UML at the design level as the source instead of using
functional analysis prototypes at the analysis level as presented in
this paper, and -- 3. extend this work to verify the refinement
process over different abstraction levels, i.e., to check if the
basic features captured at the vehicle level are correctly refined
(realized) in related functions at the analysis level as well as at
the design level where the functionalities depicted at the analysis
level are further decomposed or restructured.
%
%

\bibliographystyle{ieeetr}
\bibliography{bibtex}

\end{document}